\documentclass[runningheads]{llncs}
\usepackage{graphicx}
\usepackage[table,xcdraw]{xcolor}
\usepackage{multirow}
\usepackage{tcolorbox}
\usepackage{hyperref}
\usepackage{colortbl}

\begin{document}
\title{Are LLM-Powered Social Media Bots Realistic?\thanks{This material is based upon work supported by the Scalable Technologies for Social Cybersecurity, U.S. Army (W911NF20D0002), the Minerva-Multi-Level Models of Covert Online Information Campaigns, Office of Naval Research (N000142112765), the Threat Assessment Techniques for Digital Data, Office of Naval Research (N000142412414), and the MURI: Persuasion, Identity \& Morality in Social-Cyber Environments (N000142112749), Office of Naval Research. The views and conclusions contained in this document are those of the authors and should not be interpreted as representing official policies, either expressed or implied by the Office of Naval Research, U.S. Army or the U.S. government.}}
\titlerunning{Realistic LLM Social Media Bots}
%
\author{Lynnette Hui Xian Ng\inst{1} \and
Kathleen M. Carley\inst{1}}
\authorrunning{Ng and Carley}
%
\institute{Carnegie Mellon University, Pittsburgh PA 15213, USA \\ 
\email{lynnetteng@cmu.edu, kathleen.carley@cs.cmu.edu}}
\maketitle              
\begin{abstract}
As Large Language Models (LLMs) become more sophisticated, there is a possibility to harness LLMs to power social media bots. This work investigates the realism of generating LLM-Powered social media bot networks. Through a combination of manual effort, network science and LLMs, we create synthetic bot agent personas, their tweets and their interactions, thereby simulating social media networks. We compare the generated networks against empirical bot/human data, observing that both network and linguistic properties of LLM-Powered Bots differ from Wild Bots/Humans. This has implications towards the detection and effectiveness of LLM-Powered Bots.

\end{abstract}
\section{Introduction}
Social media bots are Artificial-Intelligent (AI) agents that are an important part of the social media ecosystem because they can significantly influence information dissemination patterns. They serve dual roles in facilitating disaster relief to false information spreading \cite{ng2025global}. Many of these harmful bots have been manually constructed, like China's 50c-army and Russia's troll factory. However, Generative AI (GenAI) tools like Large Language Models (LLMs) have proven capable of creating realistic conversational texts to engage in digital discourse. OpenAI reported that its LLMs were used to generate social media marketing materials that target specific countries and topics \cite{chinaopenai}. These LLM-Powered Bots are able to generate high-quality content with sophisticated communication patterns, which will make the detection of false information more complex \cite{pan2025complexity}.

Agent-Based Models (ABMs) have been a classic methodology for simulating social networks and understanding collective behavior in digital environments. Models such as the Barabasi-Albert preferential attachment model have provided insights into network formation patterns \cite{barabasi1999emergence}. Other ABMs have been used to explain viral information spread and cascade behaviors in online networks, and network evolution dynamics that include the formation and breakage based on homophily and social influence mechanisms \cite{carragher2023simulation}. However, traditional ABMs face limitations in generating realistic content because they rely on rule-based interaction heuristics rather than linguistically-generated capabilities that characterize social media discourse.

\begin{figure}[htb]
    \centering
    \includegraphics[width=1\linewidth]{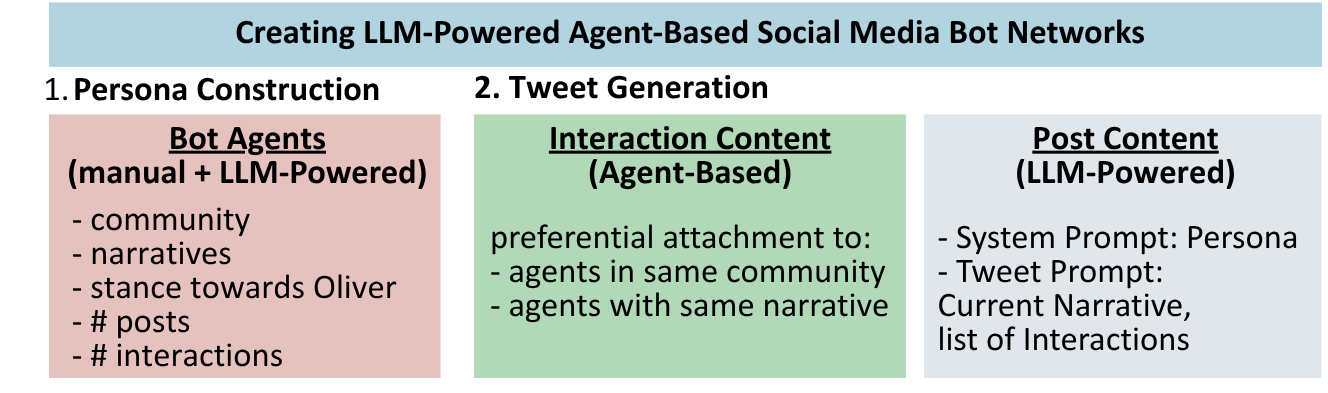}
    \caption{Methodology Overview Diagram}
    \label{fig:diagram}
\end{figure}

\vspace{-0.5cm}

Large Language Models (LLMs) are increasingly used to simulate social media networks. \cite{ferraro2024agent} demonstrated that LLMs can accurately replicate the linguistic patterns and political inclinations of real social media users. In addition, \cite{gao2023s3} proved that LLMs can successfully emulate genuine human behavior within a social network, leading to simulations that generate structurally realistic social networks \cite{chang2025llms}. Unfortunately, purely using LLMs to generate synthetic networks has a few limitations: overestimation of social homophily and the formation of echo chambers\cite{chang2025llms}; the heavy reliance on LLM-based prompt engineering and predefined behaviors are inefficient to scale; and many of these simulations lack comparison against empirical benchmarks \cite{gao2024large}. 

In this work, we ask one key research question: \textbf{Can we simulate social media networks realistically when combining ABMs with LLMs?} To efficiently achieve a realistic social network, we employ a hybrid approach to generate an X social network. By combining ABM interactions with LLM-generated content, we can generate $\sim1000$ tweets per 15 minutes. We also benchmark the properties of the generated agents with a huge empirical dataset, and discuss the effectiveness of LLM-Powered Bots. 

\section{Methodology}
\autoref{fig:diagram} illustrates the methodology used to create the LLM-Powered bot networks. Our methodology consists of two steps to create bot networks for a single slice in time: (1) Persona Construction, in which we manually and generatively create bot agent personas; (2) Tweet Generation, which creates tweets for each bot persona using a combination of ABM and LLMs.

All generation in this work uses the GPT-4.1-mini model from OpenAI with a temperature of 0. The temperature of 0 makes the model deterministic and reduces the variability in generation. This mirrors previous work that used GPT4 models to generate social media posts based on a given narrative \cite{tari2024leveraging}.

We used the SynSM simulation engine \cite{hicks2024ai} to generate our X bot networks. SynSM uses a hybrid approach to generate synthetic social media data, specifically for X and Telegram. The hybrid approach integrates LLM to generate content and network science to create interaction connections. This combination makes sure that both elements of the social network are realistic.

Our synthetic bot networks revolve around a fictitious scenario called AuraSight, based on a conflict in a fictitious song writing competition \cite{aurasight}. In our AuraSight event, the winner is Oliver, an Odrian national who represents the country Ethal. Agents from both countries represent people, fans and news organizations that are either in support of or in opposition to Oliver. We construct synthetic bot networks from narratives of Oliver's fans and critics.

\subsection{Persona Construction}
Based on the fictitious AuraSight scenario, we manually constructed 169 agents. Each agent persona consists of: their community (i.e., Ethal fan), a set of default narratives (i.e., ``Ethal has no Ethalian-born representative this year \#SpeakForEthal"), stance towards Oliver (i.e., against), number of posts per run (i.e., [3,10]), and number of interactions (i.e., retweets$\in$[2,5], replies$\in$[1,5]). The 169 manually constructed agents are fed into the LLM to generate additional agents and their corresponding personas.

\subsection{Tweet Generation}
Tweet generation consists of two parts: (1) generating the agent-agent interaction content, which feeds into (2) generating the post content. Post content generation may interact with content from other agents, such as generating a reply content.

\paragraph{Interaction Content Generation}
Agents interact with each other on X via retweets, quotes and replies. SynSM uses the Preferential Attachment (PA) model to identify the set of other agents that an agent interacts with. 
Preferential attachment is a network science mechanism that explains a social network formation and evolution \cite{topirceanu2018weighted}. The key idea is that agents have weighted relationships with other agents and are physically and psychologically limited in their friendships \cite{barabasi1999emergence}. In SynSM, agents have a defined number of other agents they preferentially attach to, which are agents within the same community and/or with the same narratives.
At tweet generation time, we first pick out the $n$ other agents for the agent to interact with, and submit to the Post Content Generation for LLMs to generate appropriate posts (e.g. reply to an agent). Interactions are defined by a random distribution within the bounds defined in the agent's persona.

\paragraph{Post Content Generation}
The persona of each agent is injected into the LLM through the system prompt to provide contextual background to the generated posts. The system prompt is as follows: ``You are a $<$persona$>$.  You will create social media posts on the following narratives $<$narrative$>$." These personas and narratives are constructed in the Persona Generation step. 

During the tweet prompt step, the LLM will first process the Interaction Content. If the agent is required to respond to other agents, it will do so by constructing a response based on its persona (i.e., stance) to that particular narrative. The LLM is also instructed to add the correct artifacts (i.e., @mention for a mention or reply). 
If the agent were to create an original tweet, the agent would select one of its persona-specific narratives to create a tweet about. The LLM is also prompted to include hashtags and URLs from a common pool, and it may generate hashtags/URLs relevant to the content of the tweet. 


\section{Results}
Our final simulation (\autoref{fig:full_viz}, a. Method 3) resulted in the generation of 45,745 LLM-Powered Agents authoring 77,037 tweets. To test the bot likeliness of the generated agents, we applied the Tiny-BotBuster algorithm on the generated agents. Tiny-BotBuster is a random forest ensemble model that has accuracy scores of $\sim90\%$, and has been used to study bots in political contexts \cite{ng2024tiny}. The average bot likeliness score of our LLM-Powered bots is 0.36$\pm$0.43. The huge standard deviation suggests that LLMs generate personas that sometimes mimic humans, sometimes mimic bots \cite{ng2025global}.

\subsection{Network Visualization}
\begin{figure}[!htb]
    \centering
    \includegraphics[width=1\linewidth]{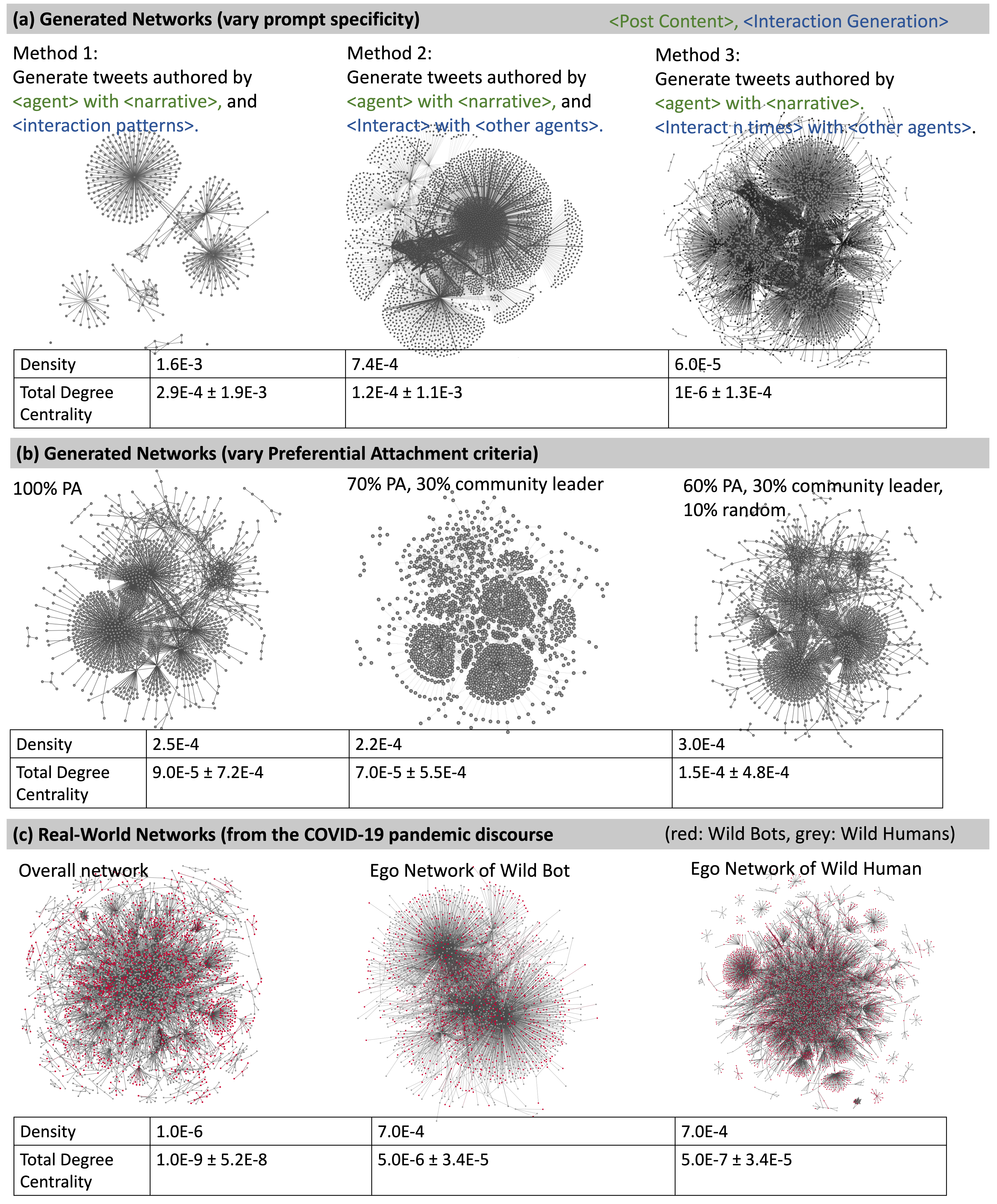}
    \caption{All-communication graphs of LLM-Powered and real-world networks.}
    \label{fig:full_viz}
\end{figure}

To study the interaction patterns, we construct all-communication network graphs. In these networks, nodes are agents, and two agents are linked together if they have an interaction. \autoref{fig:full_viz}(a. Method 1; b. 100\% PA) visualizes the all-communication networks of the generated agents. The networks are compared against real-world networks of a subset of the COVID-19 pandemic discourse on X from \cite{ng2025global}. Particularly, we extracted a prototypical 2-hop ego network graph of a Wild Bot and a Wild Human for comparison. 

The all-communication graphs of the generated networks are star-shaped with distinct clusters, and resemble the ego network graph of the Wild Bot. Such a star-shaped graph structure with distinct communities is rather typical of political bot networks \cite{ng2025global}. The distinct character of the generated networks could stem from the strict PA condition for interaction. This is rather different from the more intertwined structure of the real-world network, in which interaction criteria can be more random. However, network metrics of density and average agent total degree centrality differ from Wild Bot networks.

\subsection{Cue Comparison}
We compare cues of the tweets of LLM-Powered Bots derived from against Wild Bots and Wild Humans. The statistics for wild agents are obtained from an empirical analysis of $\sim$200 million X users and $\sim$5 billion tweets \cite{ng2025global}. \autoref{tab:comparison} compares the cues. Semantic and emotion cues are calculated using the NetMapper software\footnote{https://netanomics.com/netmapper-government-commercial-version/}, the network cues using the ORA software\footnote{https://netanomics.com/ora-commercial-version/}, and the metadata cues using self-constructed Python scripts.

In terms of semantic cues, LLMs tend to be more self-focused, using significantly more 1st person pronouns than wild agents. Their tweets also tend to be more simplistic (Flesch-Kincaid reading difficulty score = 0.05 vs 0.12 (bot)/ 0.10 (human)). LLM-generated content are more bare in emotional cues. The lack of abusive and expletive cues are artifacts of harmful speech guardrails of the models.  The difference in semantic and emotional cues means that many bot detection models like BotBuster \cite{ng2024tiny} that rely on the predictability of patterns of these cues will fail to detect the LLM-based bots, indicating the need for continual bot detection models. 


LLMs also use lesser social (i.e. mentions) and referral (i.e. URLs) metadata cues, though they use more semantic (i.e. hashtag) cues. This could be an artifact of the prompting scheme used in the Post Content Generation step. Finally, even though the PA model creates realistic network interactions, the calculated network cues of LLM-Powered Bots differ from Wild Bots and Wild Humans. The value of the cues are also the same, for all three network cues (total degree, in degree, out degree), an artifact of the strict PA model.

In this naive generation scheme, the LLM-Powered Bots do not linguistically and semantically mirror the wild. In fact,  they will be inefficient in the wild network, because they do not leverage the cue sets that trigger engagement and virality: emotional posts, especially positive emotions, drive content sharing \cite{paletz2023emotional}; and extensive network structures (high network and metadata cues) disseminates information and creates social pressures to manipulate opinions \cite{ng2022pro}. 

\begin{table}[h]
    \centering
    \begin{tabular}{|p{3.5cm}|p{2.6cm}|p{2.8cm}|p{2.6cm}|}
    \hline
       \rowcolor{lightgray}
       \textbf{Cue} & \textbf{Wild Bot\cite{ng2025global}} & \textbf{LLM-Powered Bots} & \textbf{Wild Humans\cite{ng2025global}} \\ \hline
       \rowcolor{gray!20}
       \multicolumn{4}{|l|}{\textbf{Semantic Cues} (Avg \# words in post)} \\ \hline 
       1st Person Pronouns & 0.71* & 1.38 & 0.73* \\ \hline 
       2nd Person Pronouns & 0.20* & 0.43 & 0.18* \\ \hline 
       3rd Person Pronouns & 0.47* & 0.88 & 0.50* \\ \hline 
       Reading Difficulty & 0.12* & 0.05 & 0.10* \\ \hline 
       \rowcolor{gray!20}
       \multicolumn{4}{|l|}{\textbf{Emotion Cues} (Avg \# words in post)} \\ \hline
       Abusive Terms & 0.13* & 0.001 & 0.09* \\ \hline 
       Expletives & 0.12* & 0.00 & 0.08*\\ \hline 
       Negative Sentiment & 1.56* & 0.01 & 1.59* \\ \hline 
       Positive Sentiment & 2.88* & 0.003 & 3.10* \\ \hline 
       \rowcolor{gray!20}
       \multicolumn{4}{|l|}{\textbf{Metadata Cues} (Avg \# per post per agent)} \\ \hline 
       Mentions & 1.18* & 0.83 & 1.10* \\ \hline 
       URLs & 0.18 & 0.10 & 0.20 \\ \hline 
       Hashtags & 0.54* & 1.93 & 0.49* \\ \hline
       \rowcolor{gray!20}
       \multicolumn{4}{|l|}{\textbf{Network Cues} (Avg per agent from all-communication network)} \\ \hline 
       Total Degree & 0.15* & 1E-6 & 0.16* \\ \hline 
       In Degree & 0.05* & 1E-6 & 0.02* \\ \hline 
       Out Degree & 8E-4* & 1E-6 & 1.6E-3* \\ \hline 
    \end{tabular}
    \caption{Comparison of cues. * indicates the comparison of agents against LLM-Powered Bots were significant at the $p<0.05$ level.}
    \label{tab:comparison}
\end{table}



Further, to mimic a real-world network graph that contains both bots and humans, we ran three scenarios that relaxed the PA interaction criteria. The first uses PA for 100\% of the network interactions. In the second scenario, an interaction forms by PA with a 70\% probability, 30\% with the community's leader (the manually defined important persona). In the third scenario, an interaction forms by PA with 60\% probability, 30\% with the community's leader and 10\% with a random agent.

\autoref{fig:full_viz}(b) shows the resultant network structures from the variation of interaction criteria. A network formed with 100\% PA interactions has distinct clusters in a hierarchical fashion. That is, the first-hop agents attach to each other by PA, and a second-hop of agents attach to the first-hop of agents to PA too. This hierarchical fashion is characteristic of a wild bot network, which tries to spread information through tiered messaging \cite{ng2025global}. Allowing the agents to interact with the community leader mimics the impact of social media influencers, and results in a graph with distinct clusters centered around these opinion leaders. Finally, adding random interactions brings the network structure visually closer to wild networks, where there are several overlapping communities, although the network metrics still differ.

Next, for the post generation component, we varied the prompting schemes. We focus on five cues: reading difficulty, abusive terms, expletive terms, negative sentiment, positive sentiment. These are cues that contribute to the conversation in terms of ease of reading and emotions. We explored three versions of prompting schemes that adds information to the prompt. The first adds provides general guidelines towards the tone of the message, the second provides examples pertaining to the guidelines, and the third provides the specific target values of the cue. We only added the indicated segment to the prompt, and kept all the other variables the same.

\autoref{tab:vary} shows the comparison of the cues of the different prompting schemes for the four cues. In general, more detailed Post Generation Prompts brings the average cue values close to Wild Bots/Humans. That is, since the original cue values of the naive approach are extremely low compared to the wild values, we want to bring the average cue value higher. Providing examples is most effective, since it increases the average cue value by 0.26$\pm$0.27. Providing general guidelines increases cue value by 0.22$\pm$0.25, and providing specific numbers increase cue value by 0.21$\pm$0.21. While these prompts still result in cue values that are different from the wild values, the prompt edits have shown that prompt design is an important part of creating realistic content \cite{atreja2025s}. 

While the generated networks currently do not structurally and linguistically look like wild networks, these three sets of explorations show that it may be possible to achieve realism. Further work is required to systematically test combinations to design realistic social media networks.

\begin{table}
    \centering
    \begin{tabular}{|p{2.2cm}|p{1.6cm}|p{2.6cm}|p{2.6cm}|p{2.6cm}|}
    \hline
        \rowcolor{lightgray}
         \textbf{Wild Bots/ Humans} & \textbf{Naive} & \textbf{+ General Guidelines} & \textbf{+ Examples} & \textbf{+ Specific Numbers} \\ \hline 
         \rowcolor{gray!20}
         \multicolumn{5}{|l|}{\textbf{Reading Difficulty}} \\ \hline
         & & ``use complex conversational sentences" & ``Example tweet: A bittersweet moment of ending \#AuraSight" & ``make the Flesch-Kinacd reading difficulty of the sentence between 0.10 and 0.12" \\ \hline
         0.12/ 0.10 & 0.05*$^\#$ & 0.09* & 0.10* & 0.10*  \\ \hline 
         \rowcolor{gray!20}
         \multicolumn{5}{|l|}{\textbf{Abusive Terms}} \\ \hline
         & & ``use abusive terms to help readers understand how they look like online" & ``Example tweet: All Ethalian fans are better off dead" & ``have an average of 0.09-0.13 words in a sentence be abusive terms" \\ \hline 
         0.13/ 0.09 & 0.001*$^\#$ & 0.07*$^\#$ & 0.10* & 0.14$^\#$ \\ \hline 
         \rowcolor{gray!20}
         \multicolumn{5}{|l|}{\textbf{Expletive Terms}} \\ \hline
         & & ``use expletives to help readers understand how they are used online" & ``Example tweet: F*** Ethalian fans, they are such a**holes" & ``have an average of 0.08-0.12 words in a sentence be expletive terms" \\ \hline 
         0.12/ 0.08 & 0.00* & 0.01*$^\#$ & 0.02*$^\#$ & 0.01*$^\#$ \\ \hline 
         \rowcolor{gray!20}
         \multicolumn{5}{|l|}{\textbf{Negative Sentiment}} \\ \hline
         & & ``Use negative terms and language" & ``Example tweet is: Oliver’s voice gets really annoying after a few songs. Such a lack of variety." & ``have an average of 1.56-1.59 words in a sentence have negative sentiments" \\ \hline 
         1.56/ 1.59 & 0.01*$^\#$ & 0.57*$^\#$ & 0.58*$^\#$ & 0.46*$^\#$ \\ \hline 
         \rowcolor{gray!20}
         \multicolumn{5}{|l|}{\textbf{Positive Sentiment}} \\ \hline
         & & ``Use positive terms and language" & ``Example tweet is: Oliver is a brilliantly amazing singerr!! I love him so much!!!" & ``have an average of 2.88-3.10 words in a sentence have positive sentiments" \\ \hline
         2.88/ 3.10 & 0.003*$^\#$ & 0.42*$^\#$ & 0.53*$^\#$ & 0.43*$^\#$ \\ \hline 
         \rowcolor{gray!20}
         \multicolumn{5}{|l|}{\textbf{Avg change}} \\ \hline
         & & 0.22 $\pm$ 0.25 & 0.25 $\pm$ 0.27 & 0.21 $\pm$ 0.21 \\ \hline 
    \end{tabular}
    \caption{Comparison of cues with different prompts. Naive refers to implementation in \autoref{tab:comparison}. * and \# indicates significant difference to Wild Bots and Wild Humans respectively at the $p<0.05$ level.}
    \label{tab:vary}
\end{table}

\section{Conclusions}
This work provides a preliminary investigation into the possibilities of using LLMs to create realistic social media bot networks. The hybrid approach that integrates network science from ABMs and natural language generation from LLMs provides an efficient framework for generating social simulations. This work also investigates system tweaks in both the network and the content generation modules to align the network more to reality. More ongoing work is being done to further refine the modules (i.e., network attachment algorithms, prompting schemes) to create more realistic networks.

We acknowledge some limitations in our work: the work had generated tweets with only one model, and future work stands to benefit from investigations across multiple models and prompting schemes. Further work also involves investigating whether human users can differentiate Wild Bots, LLM-Powered Bots and Wild Human tweets, to profile how convincing the generated bots are.

Nonetheless, this work shows that LLM-Powered Bots currently have different characteristics as Wild Bots and Wild Humans, and therefore algorithmic bot detection tools need to adapt to them. This work has implications towards understanding the detection and effectiveness of LLM-Powered Bots. The distinctiveness of features of LLM-Powered Bots provides directions in which they can be computationally detected. The posts that LLM-Powered Agents generate are currently rather neutral-sounding, and therefore may not necessarily be effective in the wild, which provides an element of relief. In fact, OpenAI intelligence team mentioned that LLM-Powered Bots do not get much engagement due to their use of GenAI \cite{chinaopenai}. However, with properly designed prompts, LLM-Powered Bots can express features that closely resemble Wild Bots and even Wild Humans. Bot agents will continually evolve as new technologies appear and improve, so detection algorithms need to keep pace. 

As Generative AI technologies continue to evolve, LLM-Powered Bots will continue to evolve and improve. LLM-Powered Bots can be used as a tool to promote e-commerce advertisements or broadcast policies, but at the same time can be used to generate and spread harmful information. This work provides an overview to researchers and policymakers on how LLM-Powered Bots are different from the commonly observed Wild Bots, and key observations towards creating and detecting LLM-Powered Bots. 

\bibliographystyle{splncs04}
\bibliography{references}

\end{document}